\documentclass[
 reprint,
superscriptaddress,
 amsmath,amssymb,
 aps,
superscriptaddress,
 amsmath,amssymb,
 aps,longbibliography,
floatfix,
]{revtex4-1}

\usepackage{graphicx}
\usepackage{dcolumn}
\usepackage{bm}
\usepackage{amsmath,amssymb,amstext,amsthm}
\usepackage{braket}
\usepackage{bbold}
\usepackage{bbm}
\usepackage{xcolor}
\usepackage{ gensymb }
\usepackage{soul}
\setstcolor{red}

\newcommand{\HE}[0]{$^3\text{He}$}

\begin{document}

\preprint{APS/123-QED}
\setlength{\abovedisplayskip}{1pt}
\title{Experimental Limit on Neutron Orbital Angular Momentum Detection Using Polarized $^3$He}

\author{D. Sarenac}
\email{dusansar@buffalo.edu}
\affiliation{Department of Physics, University at Buffalo, State University of New York, Buffalo, New York 14260, USA}

\author{O. Lailey} 
\affiliation{Institute for Quantum Computing, University of Waterloo,  Waterloo, ON, Canada, N2L3G1}
\affiliation{Department of Physics and Astronomy, University of Waterloo, Waterloo, ON, Canada, N2L3G1}

\author{D. V. Garrad}
\affiliation{Department of Physics and Astronomy, University of Waterloo, Waterloo, ON, Canada, N2L3G1}

\author{P. R. Vadnere}
\affiliation{Department of Physics, University at Buffalo, State University of New York, Buffalo, New York 14260, USA}

\author{N. Shentevski}
\affiliation{Department of Physics, University at Buffalo, State University of New York, Buffalo, New York 14260, USA}

\author{C. W. Clark}
\affiliation{Joint Quantum Institute, National Institute of Standards and Technology and University of Maryland, College Park, Maryland 20742, USA}

\author{D. G. Cory}
\affiliation{Institute for Quantum Computing, University of Waterloo,  Waterloo, ON, Canada, N2L3G1}
\affiliation{Department of Chemistry, University of Waterloo, Waterloo, ON, Canada, N2L3G1}

\author{J. P. Cotter}
\affiliation{Centre for Cold Matter, Blackett Laboratory, Imperial College London, Prince Consort Road, London SW7 2AZ, United Kingdom}

\author{H. Ekinci} 
\affiliation{Institute for Quantum Computing, University of Waterloo,  Waterloo, ON, Canada, N2L3G1}
\affiliation{Department of Physics and Astronomy, University of Waterloo, Waterloo, ON, Canada, N2L3G1}

\author{M. G. Huber}
\affiliation{National Institute of Standards and Technology, Gaithersburg, Maryland 20899, USA}

\author{J. W. Paster}
\affiliation{National Institute of Standards and Technology, Gaithersburg, Maryland 20899, USA}

\author{Y. Tzeng}
\affiliation{Department of Material Research, National Atomic Research Institute,  Taoyuan City 325207, Taiwan}

\author{D. Alba Venero}
\affiliation{ISIS Neutron and Muon Facility, Rutherford Appleton Laboratory, Didcot, OX110QX, United Kingdom}

\author{D. A. Pushin}
\email{dmitry.pushin@uwaterloo.ca}
\affiliation{Institute for Quantum Computing, University of Waterloo,  Waterloo, ON, Canada, N2L3G1}
\affiliation{Department of Physics and Astronomy, University of Waterloo, Waterloo, ON, Canada, N2L3G1}

\date{\today}

\pacs{Valid PACS appear here}

\begin{abstract}
A recent proposal suggested that neutron orbital angular momentum (OAM) states could be detected via spin-polarized absorption in polarized $^3$He, with predicted cross-section variations linked to the neutron's OAM. We experimentally tested this hypothesis using spin-polarized neutron beams with OAM $=-2$ to  $2$, generated by fork-dislocation phase-gratings, and transmitted through a polarized $^3$He cell. Within statistical precision, no OAM-dependent change in the absorption cross section was observed. This null result places stringent constraints on polarized $^3$He–based OAM detection schemes. The absence of an effect in the given regime is traced to the proposal's disregard of the spatial character of neutron OAM: unlike spin, OAM arises from the transverse phase structure of the wavefunction and couples only through spatial gradients and overlap. The transverse extent of neutron OAM modes expands rapidly, producing a doughnut-shaped intensity profile with negligible overlap with on-axis $^3$He nuclei, while off-axis capture samples only a locally uniform phase and reduces the interaction to the known spin dependence. These results clarify the limits of absorptive nuclear methods for probing neutron OAM and emphasize the necessity of spatially resolved interactions in any viable detection scheme.
\end{abstract}

\maketitle

\section{Introduction}

Structured quantum wavefunctions have emerged as powerful tools for probing symmetry, topology, and coherence across diverse quantum systems, from optics and electron microscopy to ultracold atoms and neutrons~\cite{rubinsztein2016roadmap,bliokh2023roadmap,ivanov2022promises}. Typical forms of structured wavefunctions include orbital angular momentum (OAM) modes~\cite{LesAllen1992,Bazhenov1990}, Bessel beams~\cite{Harris2015}, and Airy beams~\cite{berry1979nonspreading,sarenac2025generation,durnin1987diffraction}, each exhibiting distinct propagation features such as helical phase fronts, non-diffracting profiles, and self-accelerating trajectories, respectively.
In neutron science, a holographic approach using an array of fork dislocation phase-gratings has enabled the preparation of neutron beams with helical wavefronts carrying quantized OAM~\cite{sarenac2022experimental,sarenac2024small}. These developments open new avenues for fundamental tests of quantum mechanics and for novel approaches to material characterization~\cite{larocque2018twisting,schmiegelow2016transfer,afanasev2021elastic,sherwin2022scattering,geerits2025measuring}. 

Building on these advances, a recent proposal by Jach and Vinson~\cite{jach2022method} hypothesized that neutron capture on polarized \( ^3\mathrm{He} \) could serve as a definitive probe of neutron OAM. Their analysis suggests that the absorption cross section becomes sensitive not only to spin polarization, as in conventional applications, but also to the orbital structure of the neutron wavefunction, enabling direct tests of angular momentum transfer in nuclear interactions involving structured quantum states.

The reaction of interest is:

\[
n + {}^3\mathrm{He} \rightarrow p + {}^3\mathrm{H} + 764\ \mathrm{keV}
\]

\noindent which serves as a key process for thermal neutron detection. The probability of absorption is strongly influenced by the total angular momentum of the neutron–helium-3 system, making the process highly sensitive to the spin orientation of both particles. In 1957 Rose derived cross sections for absorption into both singlet and triplet spin configurations and predicted a strong sensitivity to the relative spin orientations~\cite{rose1995elementary}. Nearly a decade later, Passell and Schermer provided experimental evidence that for polarized \( ^3\mathrm{He} \) neutron absorption proceeds entirely through the singlet state~\cite{passell1966measurement}. Jach and Vinson extended this framework by predicting that when the incident neutron also carries OAM, the absorption process accesses compound nuclear states with odd parity, opening additional angular momentum channels beyond those available in spin-only interactions.

\begin{figure*}
    \centering
    \includegraphics[width=1\linewidth]{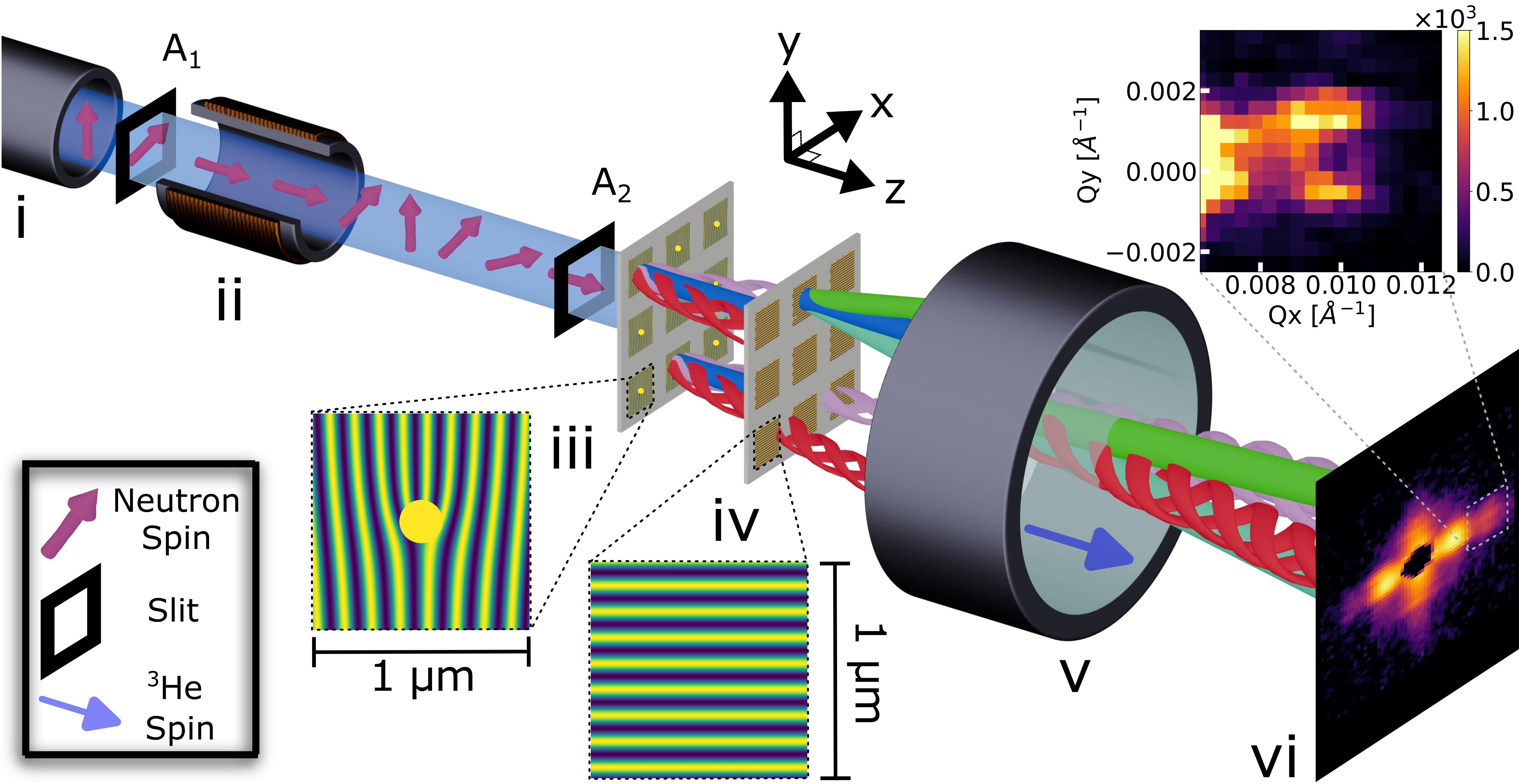}
    \caption{A schematic of the experimental setup. \textbf{(i)} The incident neutron beam is spin-polarized perpendicular to the beam propagation direction ($\pm y$) using a supermirror. Neutrons pass through the first collimating slit $A_1$ and then through the Drabkin spin flipper \textbf{(ii)}. Next, stray magnetic fields are used to rotate the spin parallel to the propagation direction ($z$). After passing through the second slit $A_2$, the beam illuminates a silicon-wafer phase grating \textbf{(iii)} with a fork-dislocation phase profile, inducing OAM in the nonzero diffraction orders. The direct beam (blue) and diffracted orders (red/pink) then pass through the $q=0$ silicon-wafer phase-grating \textbf{(iv)}, oriented perpendicular to the $q\neq0$ grating. The green/cyan beams indicate the $\pm1$ orders from the $q=0$ grating. The neutron beams then pass through a spin-polarized \HE{} cell \textbf{(v)} polarized along the beam axis, which serves as the spin analyzer. Neutrons propagate $z=10$~m to the position sensitive neutron detector. Shown in \textbf{(vi)} is the experimental intensity map (log-scale) for the multi-day measurement of $q=0$, $q=2$, integrated from $9.0$~\AA~to~$14.0$~\AA. The inset demonstrates the characteristic neutron OAM doughnut intensity profile for the $m=+2$ diffraction order from the $q=2$ grating. To resolve the OAM profile we limit the  wavelength range to $12.0$~\AA~to $13.0$~\AA~and decrease the slit sizes to reduce the beam divergence.}
    \label{fig:schematic}
\end{figure*}

Here, we report a direct experimental test of this proposal using polarized neutrons prepared in OAM states $\ell = -2,-1,0, 1, 2$, transmitted through a polarized $^3$He cell. Contrary to the theoretical expectations of Ref.~\cite{jach2022method}, we observe no measurable dependence of the absorption cross section on the neutron's OAM state. We identify the source of this discrepancy as a conceptual misinterpretation in Ref.~\cite{jach2022method}: the neutron's OAM was effectively treated as a point-like quantum number analogous to spin, without the consideration of its spatial extent. While the OAM of a helical wave can indeed be considered an intrinsic property of the state that is independent of the choice of coordinate origin~\cite{o2002intrinsic}, it is still derived from the extended transverse phase profile of the wavefunction. As such, OAM can only couple to a target system through spatial gradients and overlap. This distinction, emphasized in other neutron OAM proposals~\cite{afanasev2021elastic,sherwin2022scattering} and confirmed in atomic experiments where OAM transfer requires sub-wavelength localization~\cite{schmiegelow2016transfer}, undermines the assumptions of Ref.~\cite{jach2022method} and places strong constraints on polarized $^3$He as a probe of neutron OAM.

\section{Methods}
\label{meth}
The experiment was conducted on the ZOOM SANS instrument at the ISIS Neutron and Muon Source using polarized cold neutrons. Standard data reduction was performed using the Mantid framework~\cite{arnold2014mantid}. The usable polarized wavelength range is from 3.5~\AA~to 14.0~\AA~with a peak flux near 4.0~\AA. A schematic of the setup is shown in Fig.~\ref{fig:schematic}, illustrating the two-grating configuration and the simultaneous measurement of OAM and reference diffraction orders. A supermirror polarizer, along with a Drabkin spin flipper~\cite{Drabkin1963}, prepared the neutron beam with spin aligned perpendicular to the propagation direction ($\pm y$). The first slit, $A_1$, sits roughly 4~m before the sample, collimating the beam from $30$~mm $\times~30$~mm to $15$~mm $\times~15$~mm. The second (sample) slit, $A_2$, was $6$~mm $\times~6$~mm, approximately 30~cm from the first grating. Stray magnetic fields are used to rotate the spin parallel to the propagation direction ($z$) at the sample.

To generate neutron OAM states, silicon phase-grating samples containing arrays of fork-dislocation phase gratings were employed, as described in Ref.~\cite{sarenac2022experimental}. The fork-dislocation phase-gratings are based on a 50\% duty-cycle phase-grating profile given by
\begin{equation}
F(x) = \frac{\alpha}{2} \text{sign} \left(\sin\left[\frac{2\pi}{p}x +q\phi\right]\right),
\end{equation}
where \textit{p} is the grating period, \textit{q} is the topological charge, and $\phi$ is the azimuthal coordinate.  The parameter $\alpha$ is the phase shift induced by the grating, defined as
\begin{math}
    \alpha = - N b_c \lambda h,
\end{math} where $Nb_c$ is the coherent scattering length density of the grating material, $\lambda$ is the neutron wavelength, and \textit{h} is the grating height. Neutrons diffracted into the $\pm m$ orders carry OAM $\ell = \pm mq$. Each array consisted of 6,250,000 individual fork-dislocation phase-gratings, arranged over a 0.5~cm $\times$ 0.5~cm area. Each individual grating measured 1~$\mu$m $\times$ 1~$\mu$m, with a grating period of 120~nm and a grating height of 6~$\mu$m. Adjacent gratings were separated by 1~$\mu$m on all sides. Three grating arrays with topological charges q = 0, 1 and 2 were used in the experiment. We analyze the $\pm1$ diffraction orders for each grating, giving access to the $\ell = -2,-1, 0, +1, +2$ states.

\begin{figure}
    \centering
    \includegraphics[width=\linewidth]{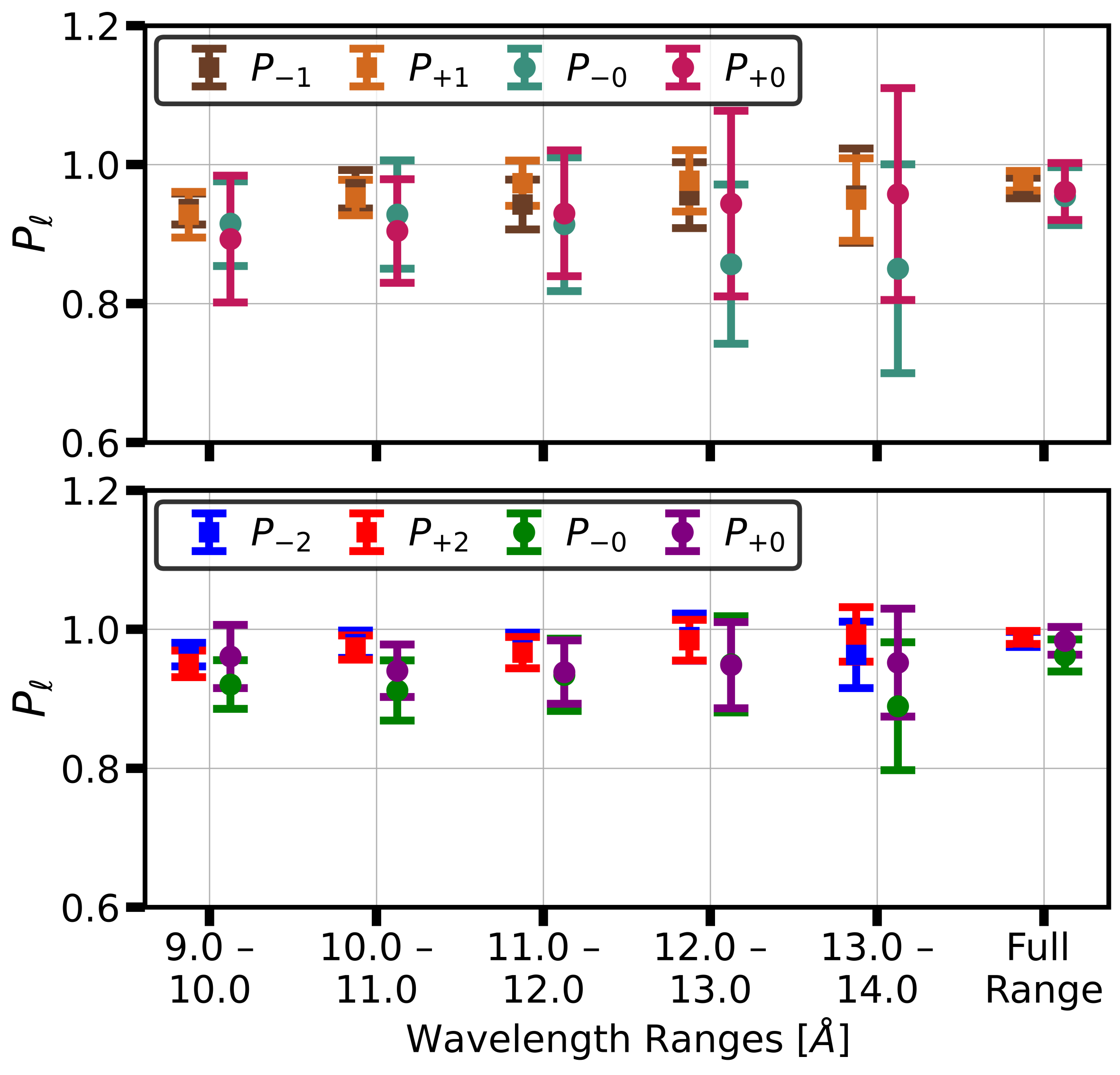}
    \caption{The measured polarization, as defined by Eq.~\ref{eqn:pol}, across the wavelength range $9.0\text{--}14.0$~\AA\ for both the $\ell = 0$ \& $1$ and $\ell = 0$ \& $2$ experimental configurations. As described in Section~\ref{meth}, the data are corrected for the depolarization of the $^3$He cell, accounting for the efficiency of the supermirror and spin flipper (see Appendix). Error bars are determined from neutron counting statistics, assuming a Poisson distribution. A potential source of systematic error arises from treating the polarization of the \HE{} cell as constant during an individual 30 minute measurement.}
    \label{fig:pdata}
\end{figure}

Unlike previous single-grating setups, this experiment employed two sequential, orthogonally oriented gratings: an extension of a method demonstrated for generating structured neutron states composed of multiple OAM modes~\cite{lailey2025multimode}. In the first configuration, a $q=0$ (linear) grating and an $q=1$ fork-dislocation grating were used; in the second configuration, a $q=0$ grating and an $q=2$ fork-dislocation grating were used. This geometry ensured that the diffraction orders from the linear grating provided an in situ reference signal, while the fork-dislocation grating simultaneously generated OAM-carrying diffraction orders. This arrangement allowed direct comparison between the OAM and reference beams under identical beam, detector, and analyzer conditions, thereby reducing systematic uncertainties associated with temporal drifts. The double grating system was mounted in an HTS magnet providing a 240~G field along the beam axis ($+z$) at the sample, to prevent neutron depolarization during transport. The gratings are mounted on a brass candle stick and loaded in from the top bore of the magnet. 

Downstream of the gratings, neutrons passed through a \HE{} cell polarized along the beam axis ($+z$), which served as a spin analyzer by preferentially absorbing neutrons with spin antiparallel to the cell polarization~\cite{gentile2017optically}. The 50~mm thick \HE{} cell was placed in a solenoid to minimize B-field gradients and maximize relaxation times. The transmitted neutrons were detected by unpolarized \HE{} detectors, and wavelength information was extracted from time-of-flight data.  

The measurements were taken across three days. The $^3$He cell was re-polarized using metastability-exchange optical pumping (MEOP) once each day~\cite{gentile2017optically}. The initial polarization of the cell was $P_0 = 0.64, 0.65, 0.63$, for the three respective cells. The relaxation times were $\tau = 37$~hours, $38$~hours, $40$~hours, and the cell pressure was $0.77$~bar, $0.69$~bar, and $0.82$~bar. The \HE{} spin direction $+z$ was kept constant during the experiment.

\begin{figure}
    \centering
    \includegraphics[width=\linewidth]{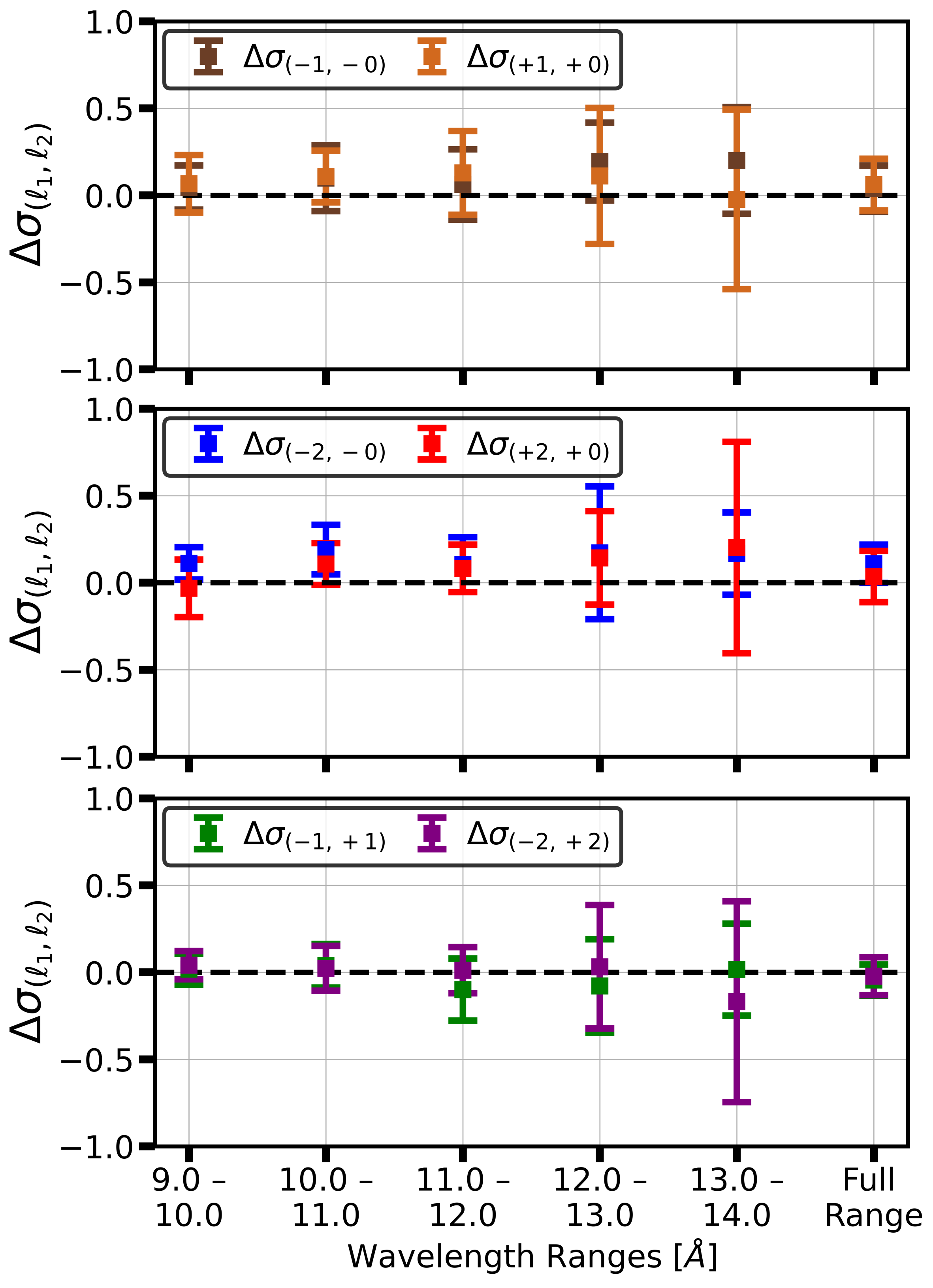}
    \caption{The normalized absorption cross section differences, as defined by Eq.~\ref{eqn:cross_diff}, across the wavelength range $9.0\text{--}14.0$~\AA\ for both the $\ell = 0$ \& $1$ and $\ell = 0$ \& $2$ experimental configurations shown in Fig.~\ref{fig:pdata}. The results indicate no statistically significant dependence of the neutron absorption cross section on the OAM state of the neutron, for both individual 1\AA~wavelength bins and the entire wavelength range. Error bars are determined from neutron counting statistics, assuming a Poisson distribution. }
    \label{fig:diffs}
\end{figure}

The measurement process was automated with a python script. The neutron spin states are prepared by flipping the neutron spin using an upstream Drabkin spin flipper. First, a 5 minute transmission measurement is taken for each spin state to monitor the polarization of the $^3$He cell. This measurement is taken in `transmission mode', which inserts a transmission monitor just inside the detector tank to monitor neutrons in the direct beam. Then, a 30 minute SANS measurement is taken for $q=0~\&~1$ or $q=0~\&~2$ gratings simultaneously at $z = 10$~m for the anti-aligned spin state ($\uparrow\downarrow$). Next, another transmission measurement is taken for each spin state, and then the final 30 minute SANS measurement for the aligned spin state ($\uparrow\uparrow$). Thus, the neutron spin pattern alternated between $\uparrow$ and $\downarrow$. Altogether, this single sequence takes roughly 1.5 hours. This measurement sequence is repeated continuously and only paused when repolarizing the \HE{} cell. In the $q=0~\&~1$ configuration, we obtained 12 complete sequences, resulting in 6 hours of SANS data per spin state. In the $q=0~\&~2$ configuration, we obtained 27 complete sequences, resulting in 13.5 hours of SANS data per spin state. Due to the depolarization of the \HE{} cell, we apply corrections to every SANS measurement based on the computed efficiency of the cell, taking into account the supermirror and RF flipper efficiency as well, before summing the loops together~\cite{arnold2014mantid, wildes2006scientific}. 

To determine the polarization of the OAM beams, we separately integrate the intensity in each of the first diffraction orders, which generate the $\pm \ell$ OAM states. The integration region is defined by a circle with radius $1.5\times10^{-3}$~\AA$^{-1}$, centered on the first diffraction orders at $(Q_x, Q_y) = (\pm5.25\times10^{-3}$~\AA$^{-1}$$,0$~\AA$^{-1}$$)$. Diffraction orders corresponding to wavelengths $<9.0$~\AA~ are not well-separated from the direct beam and excluded. This analysis is performed for both the \HE{}/neutron spin-aligned configuration ($\uparrow\uparrow$) and the \HE{}/neutron spin--anti-aligned configuration ($\uparrow\downarrow$). The resulting polarization of the individual diffraction order beams is then given by:

\begin{equation}
    P_\ell=\frac{I_{\ell, \uparrow\uparrow} - I_{\ell, \uparrow\downarrow}}{I_{\ell, \uparrow\uparrow} + I_{\ell, \uparrow\downarrow}}\label{eqn:pol},
\end{equation}

\noindent where $I_{\ell, \uparrow\uparrow}$ and $I_{\ell, \uparrow\downarrow}$ denote the intensity from the spin aligned or spin anti-aligned state respectively for OAM mode $\ell$. The polarization of the neutron OAM beams is presented in Fig.~\ref{fig:pdata}.

For a given diffraction order with OAM number $\ell$, the transmitted intensities through the polarized $^3$He cell are
$I_{\ell,\uparrow\uparrow/\uparrow\downarrow}=I_{\ell,\uparrow\uparrow/\uparrow\downarrow}^{(0)}\exp[-n\sigma_{\ell,\uparrow\uparrow/\uparrow\downarrow}L]$, where $\sigma_{\ell,\uparrow\uparrow/\uparrow\downarrow}$ are the absorption cross
sections for neutron spin parallel ($\uparrow\uparrow$) or antiparallel ($\uparrow\downarrow$) to the $^3$He polarization, $n$ is the \HE{} number density (at room temperature and specified pressure), and $L$ is the length of the \HE{} cell.
The measured polarization (Eq.\ref{eqn:pol})
therefore depends only on the spin-differential optical depth
$\Delta\tau_\ell=nL\delta\sigma_\ell$, where $\delta\sigma_\ell = \sigma_{\ell,\uparrow\downarrow}-\sigma_{\ell,\uparrow\uparrow}$. The normalized absorption cross section difference between OAM numbers $\ell_1$ and $\ell_2$ is then given by:

\begin{equation}
    \Delta\sigma_{(\ell_1, \ell_2)} =\frac{\delta\sigma_{\ell_1} - \delta\sigma_{\ell_2}}{\delta\sigma_{\ell_1} + \delta\sigma_{\ell_2}}\label{eqn:cross_diff}.
\end{equation}

Any OAM-dependent, spin-selective modification of the absorption cross section of the type predicted in
Ref.~\cite{jach2022method} would manifest as a shift
in $P_\ell$ and hence as a nonzero $\Delta\sigma_{(\ell_1,\ell_2)}$ (Eq.~\ref{eqn:cross_diff}) when comparing neutron OAM states. 

The absorption cross section differences $\Delta\sigma_{(\ell_1, \ell_2)}$ are visualized in Fig.~\ref{fig:diffs}, and summarized in Table.~\ref{tab:cross_table}. Both the $\pm1$ diffraction orders show no statistically significant deviation from zero when integrated over all wavelengths (see Table~\ref{tab:stats_table}); higher diffraction orders have insufficient counting statistics.

\begin{table*}[htbp]
  \centering
  \includegraphics[width=\textwidth]{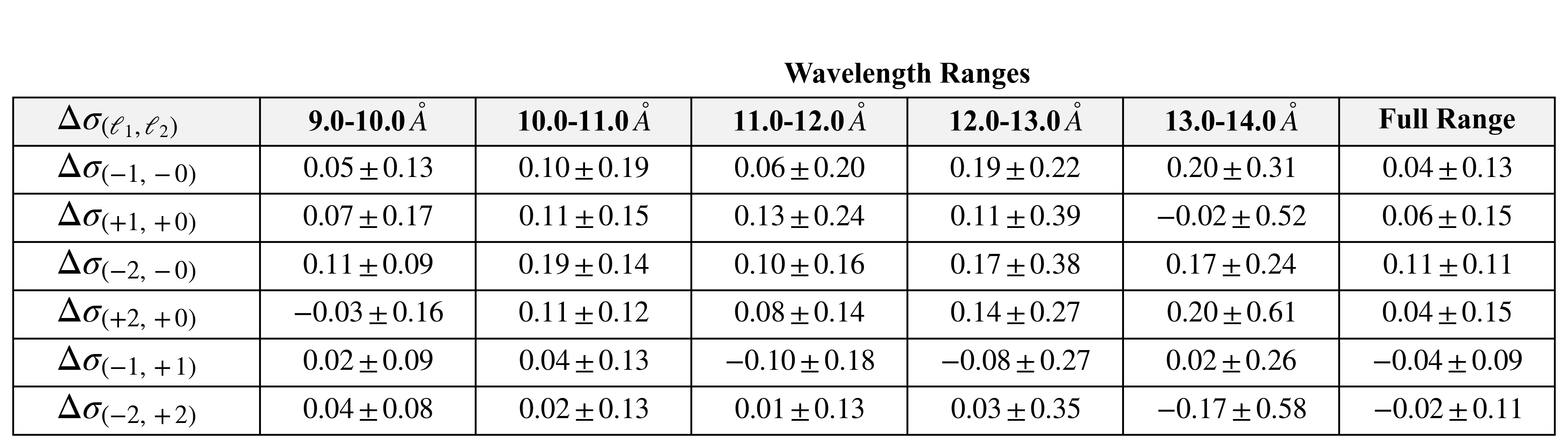}
\caption{Summary of the normalized absorption cross section differences (Eq.\ref{eqn:cross_diff}) presented in Fig.~\ref{fig:diffs}. All results are in statistical agreement with the null hypothesis $\Delta\sigma_{(\ell_1, \ell_2)} = 0$. Statistical significance is summarized in Table.~\ref{tab:stats_table}.}
  \label{tab:cross_table}
\end{table*}

\begin{table}[htbp]
  \centering
  \includegraphics[width=\linewidth]{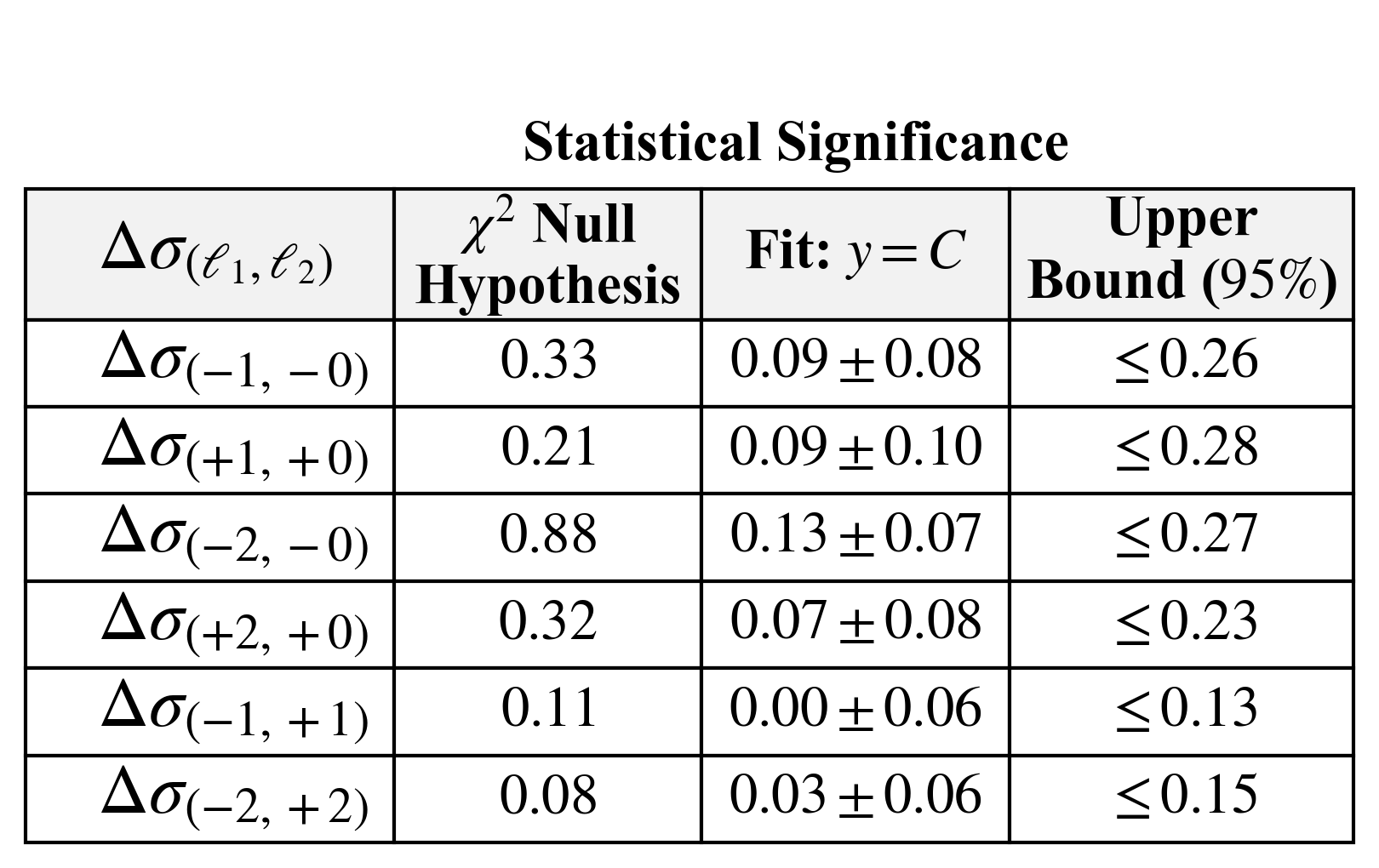}
  \caption{Statistical significance of the normalized absorption cross section differences (Eq.\ref{eqn:cross_diff}). The reduced $\chi^2$ value is computed in column 2 for the null hypothesis $\Delta\sigma_{(\ell_1, \ell_2)} = 0$. Results are $\leq1$, indicating the measurements statistically agree with the hypothesis, albeit with overestimated/conservative error bars. We fit the results to $\Delta\sigma_{(\ell_1, \ell_2)} = C$, where $C$ is a constant, to place an upper bound on the measured effect with $(95\%)$ confidence.}
  \label{tab:stats_table}
\end{table}

\section{Results and Discussion}
The results of our experiment show no statistically significant dependence of the neutron absorption cross section on the OAM state of the incoming wavefunction. Despite theoretical predictions of OAM-dependent cross sections in the polarized $^3$He($n$,$p$)$^3$H reaction~\cite{jach2022method}, our measurements with $\ell = -2,-1,0, 1, 2$ states revealed no observable variation beyond statistical uncertainty. This null result places quantitative limits on the sensitivity of polarized $^3$He absorption to neutron OAM (see Table~\ref{tab:stats_table}).

The absence of an effect can be understood by considering the spatial character of OAM. While the OAM of a helical neutron wave is a well-defined intrinsic property, conserved and independent of the coordinate origin~\cite{o2002intrinsic}, it originates from the extended transverse phase profile of the wavefunction rather than from a localized internal degree of freedom like spin. Under free propagation, the transverse structure of an OAM mode expands linearly with distance. In particular, the radius of maximum intensity of the doughnut-shaped profile scales as
$r_{\mathrm{max}} \propto \ell z \theta$,
where $z$ is the propagation distance from the grating, and $\theta$ is the beam divergence. As a result, the characteristic transverse scale of the OAM mode grows continuously from micrometer scales near the grating to millimeter scale dimensions at the position of the $^3$He cell, depending on the sample location. For OAM to influence a nuclear reaction, the target must resolve this transverse phase structure, which a polarized $^3$He gas cannot do.

At room temperature and 1 atm pressure, the mean free path of $^3$He atoms is approximately $0.14~\mu$m. Owing to the doughnut-shaped intensity profile of OAM modes, capture by on-axis $^3$He nuclei is strongly suppressed by the vanishing neutron intensity near the beam center. Absorption therefore occurs predominantly at off-axis locations, where each $^3$He nucleus samples only the local value of the neutron wavefunction. A typical neutron traverses a $^3$He cell of length $L = 5$ cm, interacting incoherently with on the order of $10^6$ different atoms. Each $^3$He nucleus therefore samples a different transverse region of the neutron wavefunction. This local sampling renders the interaction insensitive to the azimuthal phase structure of the OAM state, leaving only the established spin-dependent absorption cross section.

This interpretation aligns with insights from atomic physics. Schmiegelow et al.~\cite{schmiegelow2016transfer} showed that the transfer of optical OAM to a trapped ion requires sub-wavelength localization and sensitivity to transverse field gradients. In the neutron--$^3$He case, neither condition is satisfied: absorption probes only the local amplitude and spin degree of freedom of the neutron wavefunction and is insensitive to its transverse phase structure. 

\section {Conclusion}
Jach and Vinson predict an order-unity normalized difference in the absorption cross section between opposite OAM helicities, arising from a spin-selective modification of the neutron--$^3$He interaction.
In our experiment, no such effect is observed across a range of OAM states; the corresponding normalized cross-section differences are consistent with zero at the percent level, excluding the predicted effect by more than an order of magnitude. Therefore, our measurements rule out spin-polarized $^3$He absorption as a viable probe of neutron OAM. The assumption that OAM can act as a spin-like intrinsic quantum number in this nuclear context is not supported by experiment. Instead, any detectable neutron OAM interaction must involve spatially resolved or strongly localized couplings capable of resolving the transverse wavefunction profile. These results place stringent constraints on absorptive nuclear methods for probing neutron OAM and provide guidance for future detection strategies.

\section{Acknowledgments}
This work was supported by the Canadian Excellence Research Chairs (CERC) program, the Natural Sciences and Engineering Research Council of Canada (NSERC) Discovery program, the NSERC Canada Graduate Scholarships programs (CGS-M and PGS-D), Collaborative Research and Training Experience (CREATE) program, the Canada  First  Research  Excellence  Fund  (CFREF), and the US Department of Energy, Office of Nuclear Physics, under Interagency Agreement 89243019SSC000025. This work was also supported by the DOE Office of Science, Office of Basic Energy Sciences, in the program “Quantum Horizons: QIS Research and Innovation for Nuclear Science” through grant DE-SC0023695. Experiments at the ISIS Neutron and Muon Source were supported by beamtime allocation RB2467005 and RB2510516 from the Science and Technology Facilities Council. The authors would like to thank Mark Devonport and Kirill Nemkovski of the polarized neutron group at ISIS for their technical expertise with the $^3$He sample environment.

\bibliography{OAM}

@article{geerits2025measuring,
  title={Measuring the angular momentum of a neutron using Earth's rotation},
  author={Geerits, Niels and Sponar, Stephan and Steffen, Kyle E and Snow, William M and Parnell, Steven R and Mauri, Giacomo and Smith, Gregory N and Dalgliesh, Robert M and de Haan, Victor},
  journal={Physical Review Research},
  volume={7},
  number={1},
  pages={013046},
  year={2025},
  publisher={APS}
}

@article{lailey2025multimode,
  title={Multimode structured neutron beams},
  author={Lailey, Owen and Sarenac, Dusan and Clark, Charles W and Cory, David G and DeBeer-Schmitt, Lisa and Ekinci, Huseyin and Garrad, Davis V and Henderson, Melissa E and Huber, Michael G and Vadnere, Priyanka and others},
  journal={arXiv preprint arXiv:2511.07662},
  year={2025}
}

@article{rubinsztein2016roadmap,
  title={Roadmap on structured light},
  author={Rubinsztein-Dunlop, Halina and Forbes, Andrew and Berry, MV and Dennis, MR and Andrews, David L and Mansuripur, Masud and Denz, Cornelia and Alpmann, Christina and Banzer, Peter and Bauer, Thomas and others},
  journal={Journal of Optics},
  volume={19},
  number={1},
  pages={013001},
  year={2016},
  publisher={IOP Publishing}
}

@article{bliokh2023roadmap,
  title={Roadmap on structured waves},
  author={Bliokh, Konstantin Y and Karimi, Ebrahim and Padgett, Miles J and Alonso, Miguel A and Dennis, Mark R and Dudley, Angela and Forbes, Andrew and Zahedpour, Sina and Hancock, Scott W and Milchberg, Howard M and others},
  journal={Journal of Optics},
  volume={25},
  number={10},
  pages={103001},
  year={2023},
  publisher={IOP Publishing}
}

@article{schmiegelow2016transfer,
  title={Transfer of optical orbital angular momentum to a bound electron},
  author={Schmiegelow, Christian T and Schulz, Jonas and Kaufmann, Henning and Ruster, Thomas and Poschinger, Ulrich G and Schmidt-Kaler, Ferdinand},
  journal={Nature Communications},
  volume={7},
  year={2016},
  publisher={Nature Research}
}

@article{Harris2015,
  title={Structured quantum waves},
  author={Harris, J{\'e}r{\'e}mie and Grillo, Vincenzo and Mafakheri, Erfan and Gazzadi, Gian Carlo and Frabboni, Stefano and Boyd, Robert W and Karimi, Ebrahim},
  journal={Nature Physics},
  volume={11},
  number={8},
  pages={629--634},
  year={2015},
  publisher={Nature Publishing Group}
}

@article{methods,
  title={Methods for preparation and detection of neutron spin-orbit states},
  author={Sarenac, D and Nsofini, J and Hincks, I and Arif, M and Clark, Charles W and Cory, DG and Huber, MG and Pushin, DA},
  journal={New Journal of Physics},
  year={2018},
  volume={20},
  pages={103012}
}

@article{larocque2018twisting,
  title={Twisting neutrons may reveal their internal structure},
  author={Larocque, Hugo and Kaminer, Ido and Grillo, Vincenzo and Boyd, Robert W and Karimi, Ebrahim},
  journal={Nature Physics},
  volume={14},
  number={1},
  pages={1},
  year={2018},
  publisher={Nature Publishing Group}
}

@article{sherwin2022scattering,
  title={Scattering of slow twisted neutrons by ortho-and parahydrogen},
  author={Sherwin, Joseph A},
  journal={Physics Letters A},
  pages={128102},
  year={2022},
  publisher={Elsevier}
}

@article{jach2022method,
  title={Method for the definitive detection of orbital angular momentum states in neutrons by spin-polarized He 3},
  author={Jach, Terrence and Vinson, John},
  journal={Physical Review C},
  volume={105},
  number={6},
  pages={L061601},
  year={2022},
  publisher={APS}
}

@article{afanasev2021elastic,
  title={Elastic scattering of twisted neutrons by nuclei},
  author={Afanasev, Anatoli V. and Karlovets, Dmitry V. and Serbo, Valeriy G.},
  journal={Physical Review C},
  volume={103},
  number={5},
  pages={054612},
  year={2021},
  publisher={APS}
}

@article{Bazhenov1990,
  title={Laser beams with wave front screw dislocations},
  author={Bazhenov, Vladimir Yu. and Vasnetsov, Mikhail V. and Soskin, Marat S.},
  journal={Pisma v Zhurnal Eksperimentalnoi i Teoreticheskoi Fiziki},
  volume={52},
  pages={1037--1039},
  year={1990}
}

@article{LesAllen1992,
  title={Orbital angular momentum of light and the transformation of Laguerre-Gaussian laser modes},
  author={Allen, Les and Beijersbergen, Marco W and Spreeuw, RJC and Woerdman, JP},
  journal={Physical review A},
  volume={45},
  number={11},
  pages={8185},
  year={1992},
  publisher={APS}
}

@article{passell1966measurement,
  title={Measurement of the Spin Dependence of the {H}e 3 (n, p) T Reaction and of the Nuclear Susceptibility of Adsorbed {H}e 3},
  author={Passell, L and Schermer, RI},
  journal={Physical Review},
  volume={150},
  number={1},
  pages={146},
  year={1966},
  publisher={APS}
}

@article{durnin1987diffraction,
  title={Diffraction-free beams},
  author={Durnin, James and Miceli, Jr., John J. and Eberly, Joseph H.},
  journal={Physical Review Letters},
  volume={58},
  number={15},
  pages={1499},
  year={1987},
  publisher={American Physical Society}
}

@book{rose1995elementary,
  title={Elementary theory of angular momentum},
  author={Rose, Morris Edgar},
  year={1995},
  publisher={Courier Corporation}
}

@article{berry1979nonspreading,
  title={Nonspreading wave packets},
  author={Berry, Michael V and Balazs, Nandor L},
  journal={American Journal of Physics},
  volume={47},
  number={3},
  pages={264--267},
  year={1979},
  publisher={American Association of Physics Teachers}
}

@article{o2002intrinsic,
  title={Intrinsic and extrinsic nature of the orbital angular momentum of a light beam},
  author={O'Neil, Anna T. and MacVicar, Ignatius and Allen, Les and Padgett, Miles J.},
  journal={Physical Review Letters},
  volume={88},
  number={5},
  pages={053601},
  year={2002},
  publisher={APS}
}

@article{oam,
  title={Controlling neutron orbital angular momentum},
  author={Clark, Charles W and Barankov, Roman and Huber, Michael G and Arif, Muhammad and Cory, David G and Pushin, Dmitry A},
  journal={Nature},
  volume={525},
  number={7570},
  pages={504--506},
  year={2015},
  publisher={Nature Publishing Group}
}

@article{ivanov2022promises,
  title={Promises and challenges of high-energy vortex states collisions},
  author={Ivanov, Igor P},
  journal={Progress in Particle and Nuclear Physics},
  pages={103987},
  year={2022},
  publisher={Elsevier}
}

@article{sarenac2022experimental,
  title={Experimental realization of neutron helical waves},
  author={Sarenac, Dusan and Henderson, Melissa E and Ekinci, Huseyin and Clark, Charles W and Cory, David G and DeBeer-Schmitt, Lisa and Huber, Michael G and Kapahi, Connor and Pushin, Dmitry A},
  journal={Science Advances},
  volume={8},
  number={46},
  pages={eadd2002},
  year={2022},
  publisher={American Association for the Advancement of Science}
}

@article{sarenac2024small,
  title={Small-angle scattering interferometry with neutron orbital angular momentum states},
  author={Sarenac, Dusan and Henderson, Melissa E and Ekinci, Huseyin and Clark, Charles W and Cory, David G and DeBeer-Schmitt, Lisa and Huber, Michael G and Lailey, Owen and White, Jonathan S and Zhernenkov, Kirill and others},
  journal={Nature Communications},
  volume={15},
  number={1},
  pages={10785},
  year={2024},
  publisher={Nature Publishing Group UK London}
}

@article{sarenac2025generation,
  title={Generation of neutron Airy beams},
  author={Sarenac, Dusan and Lailey, Owen and Henderson, Melissa E and Ekinci, Huseyin and Clark, Charles W and Cory, David G and DeBeer-Schmitt, Lisa and Huber, Michael G and White, Jonathan S and Zhernenkov, Kirill and others},
  journal={Physical Review Letters},
  volume={134},
  number={15},
  pages={153401},
  year={2025},
  publisher={APS}
}

@article{wildes2006scientific,
  title={Scientific reviews: Neutron polarization analysis corrections made easy},
  author={Wildes, Andrew R.},
  journal={Neutron News},
  volume={17},
  number={2},
  pages={17--25},
  year={2006},
  publisher={Taylor \& Francis}
}

@article{gentile2017optically,
  title={Optically polarized {H}e 3},
  author={Gentile, Thomas R and Nacher, PJ and Saam, B and Walker, TG},
  journal={Reviews of modern physics},
  volume={89},
  number={4},
  pages={045004},
  year={2017},
  publisher={APS}
}

@article{arnold2014mantid,
  title={Mantid—Data analysis and visualization package for neutron scattering and $\mu$ SR experiments},
  author={Arnold, Owen and Bilheux, Jean-Christophe and Borreguero, JM and Buts, Alex and Campbell, Stuart I and Chapon, L and Doucet, Mathieu and Draper, N and Leal, R Ferraz and Gigg, MA and others},
  journal={Nuclear instruments and methods in physics research section a: accelerators, spectrometers, detectors and associated equipment},
  volume={764},
  pages={156--166},
  year={2014},
  publisher={Elsevier}
}

@article{Drabkin1963,
  author  = {Drabkin, G. M.},
  title   = {Resonance neutron spin flipper},
  journal = {Soviet Physics JETP},
  volume  = {16},
  pages   = {781--786},
  year    = {1963}
}

\section*{Appendix}
\begin{figure}[h]
    \centering
    \includegraphics[width=\linewidth]{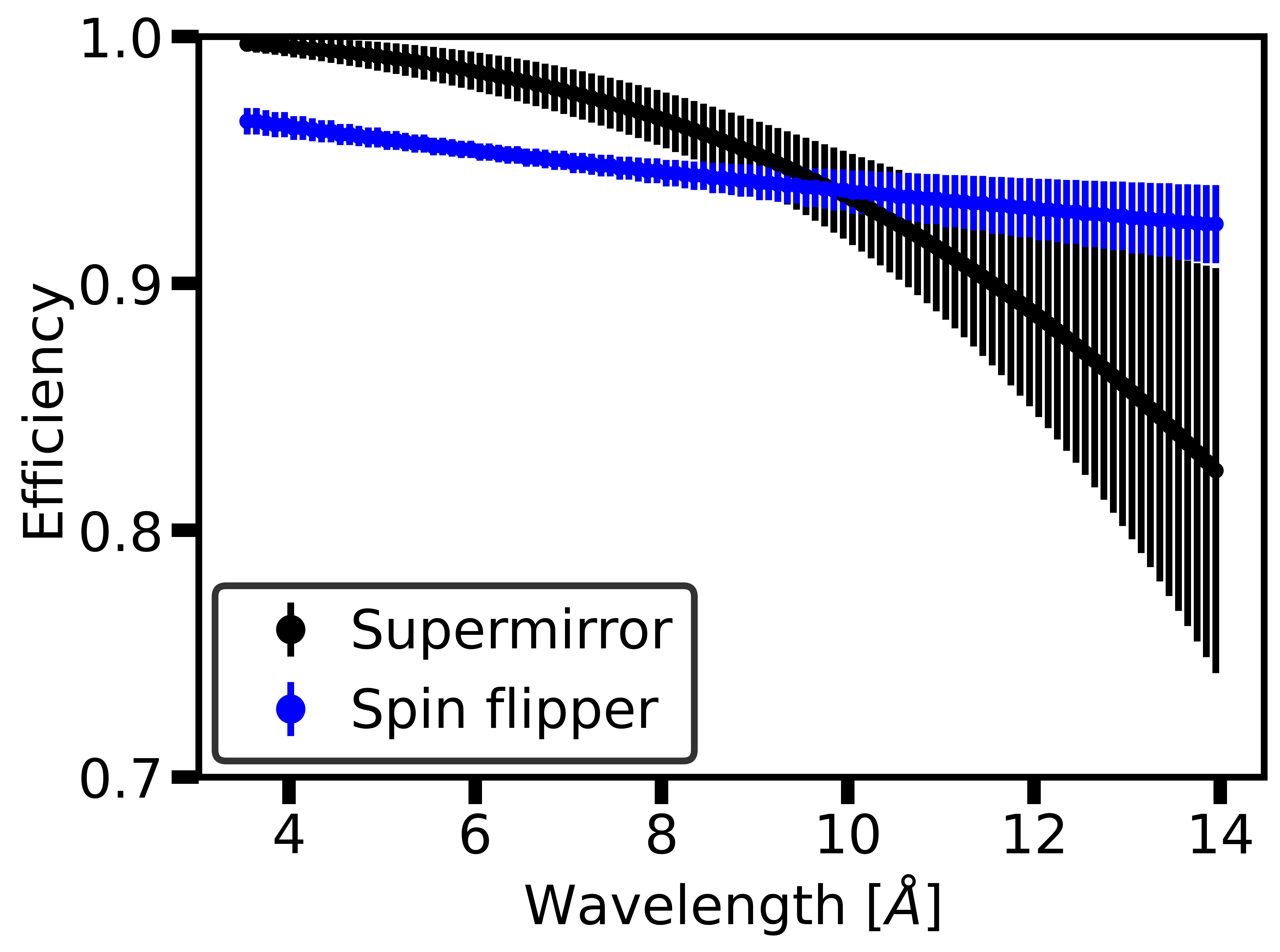}
    \caption{Efficiency of the supermirror and Drabkin spin flipper at ZOOM for polarizing the neutron beam and selecting the $\uparrow$, $\downarrow$ spin states.}
    \label{fig:effic}
\end{figure}

The efficiencies of the supermirror and spin flipper used for polarization correction of the raw data are shown in Fig.~\ref{fig:effic}.

\end{document}